\documentclass[epj]{svjour}
\usepackage{graphics}
\begin{document}
\title{The LHC state at 125.5 GeV and FNAL data as
  an evidence for 
the existence of the new class of particles -- $W$-hadrons}
%\subtitle{Bogoliubov compensation principle in the EW interaction}
\author{Boris A. Arbuzov}%\inst{1} \and Second author\inst{2}% etc

    % Insert a name or remove this line
%
\institute{Skobeltsyn Institute of Nuclear Physics of
MSU, 119991 Moscow, Russia}
\date{Received: date }%/ Revised version: date}
\abstract{
The recently discovered resonance at $125.5\, GeV$ in  invariant mass distribution of $\gamma\, \gamma$ and of $l^+\,l^+\,l^-\,l^-$ may be  tentatively interpreted as a scalar bound state $X$ consisting of two $W$.  In the present note we consider this option and show that this interpretation agrees existing experimental data including the last LHC discovery and the $b \bar b$ bump reported by CDF and D0 collaborations at TEVATRON.  The application of this scheme gives satisfactory agreement with existing data without any adjusting parameter but the bound state mass $125.5\,GeV$. There are pronounced distinctions of the $W$-hadron option from the SM Higgs case in decay mode $\,X \to \gamma\, l^+ l^-$ and in the cross-section of process $p + p \to \gamma\, X$.}

%\keywords{anomalous three-boson
 %interaction; the Higgs boson search; W-hadrons}
\PACS{{12.15.-y}{}   \and
     {14.70.Fm}{}   \and
     {14.80.Ec}{}}
\authorrunning{B.A. Arbuzov}
\titlerunning{The LHC 125.5 GeV state as a $W$-hadron}
     %} % end of PACS codes
%end of abstract
%Shorttitle:{CMS ridge effect and bremstralung of bosons off accelerated quarks } 
\maketitle
 
%\section{Introduction}
%\label{intro}
\section{Strong effective three-boson interaction}

Recent LHC searches for Higgs scalar~\cite{LHC1,LHC2,LHC3,LHC4} result in the outstanding discovery of a state with mass around $125\, GeV,$ which manifest itself in decays to $\gamma\,\gamma$ and $l^+l^+l^-l^-$. The data are consistent with the production of the SM Higgs scalar. However in numerous comments the results are considered not only in terms of the SM Higgs, but also in different extensions of the SM. In any case data being presented in~\cite{LHC1,LHC2,LHC3,LHC4} allow discussion of different options the more so, as the agreement of the data with SM predictions is not very convincing.

The present note is based mostly on works~\cite{AZPR,AZ12}.
We would discuss an interpretation of the LHC $125.5\,GeV$ state in terms of non-perturbative effects of the electro-weak interaction. For the purpose we rely on an approach induced by N.N. Bogoliubov compensation principle~\cite{Bog1,Bog2}.
In works~\cite{BAA04}- \cite{AZ11}, this approach
was applied to studies of a spontaneous generation of effective non-local interactions in renormalizable gauge theories.
In particular, papers~\cite{BAA09,AZ11}  deal with an application of the approach to the electro-weak interaction and a possibility of spontaneous generation of effective anomalous three-boson interaction of the form
\begin{eqnarray}
& &-\,\frac{G}{3!}\,F\,\epsilon_{abc}\,W_{\mu\nu}^a\,W_{\nu\rho}^b\,W_{\rho\mu}^c\,;
\nonumber\\
& &W^3_{\mu \nu}\,=\,\cos\theta_W\,Z_{\mu \nu}\,+\,\sin\theta_W\,A_{\mu \nu}\,;\label{FFF}\\
& &W_{\mu\nu}^a\,=\,
\partial_\mu W_\nu^a - \partial_\nu W_\mu^a\,+g\,\epsilon_{abc}W_\mu^b W_\nu^c\,.\nonumber
\end{eqnarray}
with uniquely defined form-factor $F(p_i)$, which guarantees effective interaction~(\ref{FFF}) acting in a limited region of the momentum space. It was done of course in the framework of an approximate scheme, which accuracy was estimated to be $\simeq 10\%$~\cite{BAA04}. 

Would-be existence of effective interaction~(\ref{FFF}) leads to important non-perturbative effects in the electro-weak interaction. It is
usually called anomalous three-boson interaction and it is considered for long time on phenomenological grounds~\cite{Hag1,Hag2}.  Our interaction constant $G$ is connected with
conventional definitions in the following way
\begin{equation}
G\,=\,-\,\frac{g\,\lambda}{M_W^2}\,;\label{Glam}
\end{equation}
where $g \simeq 0.65$ is the electro-weak coupling.
The current limitations for parameter $\lambda$ read~\cite{EW,D01}
\begin{equation}
-\,0.059< \lambda < 0.026\,;\quad 
-0.036\,<\,\lambda\,<\,0.044\,;\;(95\%\,C.L.)\,.\label{lambda1}
\end{equation}

Interaction~(\ref{FFF}) increases with increasing momenta $p$. For estimation of an effective dimensionless coupling we choose symmetric momenta (p\,,q\,,k) in vertex corresponding to the interaction
\begin{eqnarray}
& &(2\pi)^4\,G\,\,\epsilon_{abc}\,(g_{\mu\nu} (q_\rho pk - p_\rho qk)+ \nonumber\\
& &g_{\nu\rho}
(k_\mu pq - q_\mu pk)+g_{\rho\mu} (p_\nu qk - k_\nu pq)+\label{vertex}\\
& &+\,q_\mu k_\nu p_\rho - k_\mu p_\nu q_\rho)\,F(p,q,k)\,
\delta(p+q+k)\,+...;\nonumber
\end{eqnarray}
where
$p,\mu, a;\;q,\nu, b;\;k,\rho, c$ are respectfully incoming momenta,
Lorentz indices and weak isotopic indices
of $W$-bosons. 
Explicit expression for the corresponding vertex is presented in work~\cite{BAA09}.
Form-factor $F(p,q,k)$ is obtained in work~\cite{AZ11} using the following approximate dependence on the three variables
\begin{equation}
F(p,q,k)\,=\,F\Bigl(\frac{p^2+q^2+k^2}{2}\Bigr).\label{FP}
\end{equation}
Symmetric condition means
\begin{equation}
pq\,=\,pk\,=\,qk\,=\,\frac{p^2}{2}\,=\,\frac{q^2}{2}\,=\,\frac{k^2}{2}\,.
\end{equation}
Interaction~(\ref{FFF}) increases with increasing momenta $p$ and
corresponds to effective dimensionless coupling being of the following order of magnitude
\begin{equation}
g_{eff}\,=\,\frac{|g\,\lambda|\,p^2}{2 M_W^2}\,F\Bigl(\frac{3\,p^2}{2}\Bigr)\,.\label{geff}
\end{equation}
Behavior of $g_{eff}(t)$ is presented at Fig. 1.

\resizebox{0.45\textwidth}{150pt}{
  \includegraphics{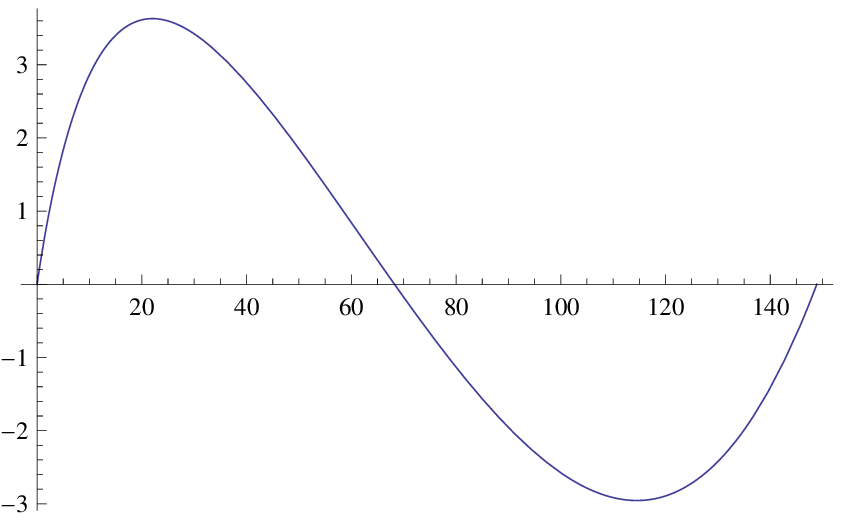}
}
%\begin{figure}[ht]
%\epsfig{file=Gg.eps,width=350pt, height=150pt}
\\
Fig. 1. Behavior of the effective coupling $g_{eff}(t),\,t=G\,p^2$; $g_{eff}(t)\,=\,0 $ for $t\,>\,148$ .
%\label{GEF}
%\end{figure}
\bigskip

We see that for $t \simeq 22$ the coupling reaches maximal value $g_{eff} \,=\, 3.63$ (e.g. $p(max)\,\simeq\,5.4\,TeV$ with $G$ from the forthcoming solution), that is corresponding effective $\alpha$ is the following
\begin{equation}
\alpha_{eff} = \frac{g_{eff}^2}{4\,\pi}\,=\,1.049\,.\label{aleff}
\end{equation}
Thus for sufficiently large momentum, interaction~(\ref{FFF})  becomes strong
and may lead to physical consequences analogous to that of the usual strong interaction (QCD). In particular bound states and resonances constituting of $W$-s (W-hadrons) may appear.
We have already discussed~\cite{AZ12} a possibility to interpret the would-be CDF $Wjj$ excess~\cite{CDF} in terms of such state.

\section{Scalar bound state of two W-s}

In the present note we apply these considerations along with some results of work~\cite{AZ11} to the discovered decays to $\gamma\,\gamma$ and $l^+\,l^+\,l^-\,l^-$ of LHC $125.5\,GeV$ state~\cite{LHC1}- \cite{LHC4}.

Let us assume that this effect is due to existence of
bound state $X$ of two $W$ with mass $M_s$. This state $X$ is assumed to have
spin 0 and weak isotopic spin also 0. Then vertex of $XWW$ interaction has the following form
\begin{equation}
\frac{G_X}{2}\,\,W_{\mu \nu}^a\,W_{\mu \nu}^a\,X\,\Psi_0\,;
\label{XWW}
\end{equation}
where $\Psi_0$ is a Bethe-Salpeter wave function of the bound state.
The main interactions forming the bound state are just non-perturbative interactions~(\ref{FFF}, \ref{XWW}). This means that we take into account exchange of vector boson $W$ as well as of scalar bound state $X$ itself. In diagram form the corresponding Bethe-Salpeter equation is presented in Fig. 2.

\resizebox{0.45\textwidth}{150pt}{
  \includegraphics{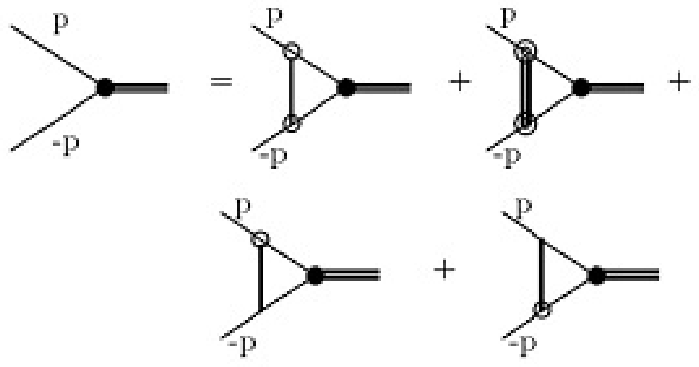}}
%\begin{figure}[ht]
%\epsfig{file=Fig2d.eps,width=350pt, height=150pt}
\\
Fig. 2. Diagram representation of Bethe-Salpeter
equation for W-W bound state. Black spot corresponds to $XWW$ vertex~(\ref{XWW}) with BS wave function. Empty circles correspond to point-like anomalous
three-boson vertex~(\ref{FFF}), double circle -- point-like XWW vertex~(\ref{XWW}). Simple point -- usual gauge triple $W$ interaction. Double line -- the bound state $X$, simple line -- W.
\bigskip
%\end{figure}

We solve equation Fig. 2 with account of normalization conditions for Bethe-Salpeter wave function (details in work~\cite{AZPR}).

We introduce $M_s\,=\,125.5\,GeV$ and then we have unique solution of the set of equations and conditions with the following parameters
\begin{equation}
 G_X = 0.000666\,GeV^{-1} ;\quad
 G = \frac{0.00484}{M_W^2}\,.\label{GGX}
\end{equation}
Result~(\ref{GGX}) means parameter of anomalous triple interaction~(\ref{FFF}) with account of relation~(\ref{Glam})
\begin{equation}
 \lambda\,=\,-\frac{G\,M_W^2}{g}\,=\,-\,0.00744\,;\label{lambda}
\end{equation}
which doubtless agrees limitations~(\ref{lambda1}).\\
\section{Experimental implications}
Thus we have scalar state $X$ with coupling~(\ref{XWW}, \ref{GGX}). In calculations of decay parameters and cross-sections we use CompHEP package~\cite{Boos}. We use parameter $G_X$~(\ref{GGX}) being obtained above and $M_s=125.5\,GeV$.
Cross-sections of $X$ production at LHC for energies being used in works\cite{LHC1}- \cite{LHC4} read
\begin{eqnarray}
& &\sigma_X = \sigma(p+p\to X+...) = 0.18 pb; \;\sqrt{s} = 7 TeV;\label{Section}\\
& &\sigma_X = \sigma(p+p\to X+...) = 0.21 pb ; \;\sqrt{s} = 8 TeV.\nonumber
\end{eqnarray}
Parameters of $X$-decay are the following
\begin{eqnarray}
& &\Gamma_t(X) = 0.000502\,GeV ;\nonumber\\
& & BR(X \to \gamma \gamma)\,=\,0.430 ;\quad BR(X \to \gamma Z)\,=\,0.305 ;\nonumber\\
& &BR(X \to 4\,l (\mu, e))\,=\,0.00092 ;\quad\nonumber\\
& & BR(X \to b\, \bar b)\,=\,  0.000024\,.\nonumber\\
& & BR(X \to \gamma e^+ e^-)\,=\,0.0231 ;\;\nonumber\\
& & BR(X \to \gamma \mu^+ \mu^-)\,=\,0.016 ;\label{decay}\\
& & BR(X \to \gamma \tau^+ \tau^-)\,=\,0.0125 ;\;\nonumber\\
& & BR(X \to \gamma u \bar u)\,=\,0.0478 ;\nonumber\\
& & BR(X \to  \gamma c \bar c)\,=\,0.0368 ;\quad BR(X \to \gamma d \bar d)\,=\,0.0446 ;\nonumber\\
& &BR(X \to \gamma s \bar s))\,=\,0.0430 ;\quad BR(X \to \gamma b \bar b)\,=\,  0.0416\,.\nonumber
\end{eqnarray}
For decay $X \to b \bar b$ we calculate the evident triangle diagram and use $m_b(125\,GeV) \simeq 2.9\,GeV$. Branching ratios for decays to
other fermion pairs are even smaller. We see that state $X$ is quite narrow, so we would expect the observable width of the state to be defined by the corresponding experimental resolution.

Experimental data give in the region of the state the following results for $\sigma_{\gamma \gamma}\,=\,\sigma_X\,BR(X \to \gamma \gamma)$~\cite{LHC3,LHC4}
\begin{eqnarray}
& &\mu_{\gamma \gamma}\,=\frac{\sigma\times BR(X\to \gamma \gamma)_{exp}}{\sigma\times BR(X\to \gamma \gamma)_{SM}}\,=\,1.8\pm 0.5\,;\label{limit}\\
& &\mu_{\gamma \gamma}\,=\frac{\sigma\times BR(X\to \gamma \gamma)_{exp}}{\sigma\times BR(X\to \gamma \gamma)_{SM}}\,=\,1.6\pm 0.4\,.\nonumber
\end{eqnarray}
Here $\sigma\times BR(X\to \gamma \gamma)_{SM}\simeq\,0.04\,pb$ is the Standard Model value for the quantity under discussion, upper line corresponds to ATLAS data~\cite{LHC3} and the lower line corresponds to CMS data~\cite{LHC4}. Firstly both limitations are quite consistent. Secondly our value for the same quantity from~(\ref{Section}, \ref{decay}) reads
\begin{equation}
\mu_{\gamma \gamma}\,=\frac{\sigma\times BR(X\to \gamma \gamma)_{calc}}{\sigma\times BR(X\to \gamma \gamma)_{SM}}\,=\,1.9\,;\label{CS0}
\end{equation}
that also agrees results~(\ref{limit}), however it essentially exceeds the SM value. At this point it is advisable to discuss accuracy of our approximations.
The former experience concerning both applications to Nambu -- Jona-Lasinio model in QCD~\cite{BAA06,AVZ06,AVZ09} and to the electro-weak interaction~\cite{BAA09,AZ11} shows that average accuracy of the method is around 10\% in values of different parameters. So we may assume, that in the present estimations of coupling constant $G_X$ we also have the same accuracy. For the cross-section this means possible deviation up to 20\% of the calculated value. Thus we would change~(\ref{CS0}) to
the following result
\begin{equation}
\mu_{\gamma \gamma}\,=\,(1.9\pm 0.38)\,pb\,;\label{CS1}
\end{equation}
Branching ratios~(\ref{decay}) do not depend on the value of $G_X$, so we assume their accuracy being considerably better than in~(\ref{CS1}).
In any case result~(\ref{CS1}) agrees~(\ref{limit}).

We would emphasize importance of channel $X \to \gamma\, l^+ l^-$. For this decay mode from~(\ref{Section}, \ref{decay}) we predict for energy $\sqrt{s} = 8\,TeV$
\begin{equation}
\sigma_X\,BR(X \to \gamma l^+ l^-)\,=\,(0.0075 \pm 15)\,pb;
\label{All}
\end{equation}
that gives $N \sim 70 $ events for already achieved luminosity~\cite{LHC3,LHC4}. This channel may serve for an accurate test of our results because the SM Higgs option gives around 5 events~\cite{Gainer}. By the way, authors of work~\cite{Gainer} call this channel "overlooked" and I incline to agree this definition, because the channel can be effectively registered but have not been studied yet.

The important difference of our predictions with the SM results consists in decay channel $X\to b \bar b$. For SM Higgs which is usually considered for explanation of $125.5\,GeV$ state this decay is dominant, whereas our result~(\ref{decay}) gives extremely small $BR\,\simeq 3\,10^{-5}$. We would emphasize that SM Higgs interpretation could not be considered as proved unless $b \bar b$ channel with the proper intensity would be detected. However recently the results of TEVATRON were reported~\cite{D0CDF}, in which there was an excess of $b \bar b$ events registered in the region $120\,GeV\,<\,M_{bb}\,<\,150$. Provided this excess being prescribed to decay of Higgs the result reads~\cite{D0CDF}
\begin{equation}
\mu_{bb}\,=\,1.97^{+0.74}_{-0.73}\,;\label{bB}
\end{equation}
that the authors of~\cite{D0CDF} consider as a confirmation of SM Higgs interpretation of results~\cite{LHC1,LHC2,LHC3,LHC4}. We shall once more discuss this item after a consideration of the vector $W\,W$ bound state.

\section{Vector isovector state and the $b \bar b$ bump at TEVATRON}

In work~\cite{AZ12} the interpretation of CDF $jet-jet$ enhancement around 140 GeV~\cite{CDF} was considered as a manifestation of isovector $W$-hadron with spin 1. 
We assume that this excess is due to existence of 
bound state $V$ of two $W$. This state $V$ is assumed to have 
spin 1 and weak isotopic spin also 1. Then vertex of $VWW$ interaction has the following form
\begin{equation}
\frac{G_V}{2}\,\epsilon_{a b c}\,W_{\mu \nu}^a\,W_{\nu \rho}^b\,V_{\rho \mu}^c\,\Psi_V\,;
\label{VWW}
\end{equation}
where $\Psi_V$ is a Bethe-Salpeter wave function of the bound state. The main interactions forming the bound state are just non-perturbative interactions~(\ref{FFF}, \ref{VWW}). This means that we take into account exchange of vector boson $W$ as well as of vector bound state $V$ itself. 
Bethe-Salpeter wave function $\Psi_V$ provides effective form-factor $F_V(p)\equiv \Psi_V(p)$. Form-factor $F_V(p)$ in work~\cite{AZ12} is expressed in terms of the Meijer functions
\begin{eqnarray}
& &F_V(p) = \frac{\pi}{2} G_{15}^{21}\Bigl( z |^0_{1, 0, 1/2, -1/2, -1}
\Bigr) + \nonumber\\
& &C_1 G_{15}^{21}\Bigl( z |^{1/2}_{1/2, 1/2,
 1, -1/2, -1}\Bigr) +\nonumber\\
& &C_2 G_{04}^{20}\Bigl( z |1/2, 1, -1/2, -1\Bigr) +\nonumber\\
& &
C_3 G_{04}^{10}\Bigl( - z |1, 1/2, -1/2, -1\Bigr).\label{solutiong}\\
& &
z = \frac{G_V^2\,(p^2)^2}{1024\,\pi^2};\; C_1 = -\,0.015282;\;\nonumber\\
& & C_2 = -\,3.6512;\; C_3 = 1.28\,10^{-11}.\nonumber
\end{eqnarray}
Here $G_V$ is taken to be 
\begin{equation}
G_V\,=\,\frac{0.1425}{M_W^2}\,.\label{G}
\end{equation}
It is 15\% smaller than the value being obtained in~\cite{AZ12}. We take this value within the accuracy of the method in view to obtain 
consistent agreement of the totality of data to experiments. Behavior of $F_V(p)$ is presented in Fig.3.

\resizebox{0.45\textwidth}{150pt}{
  \includegraphics{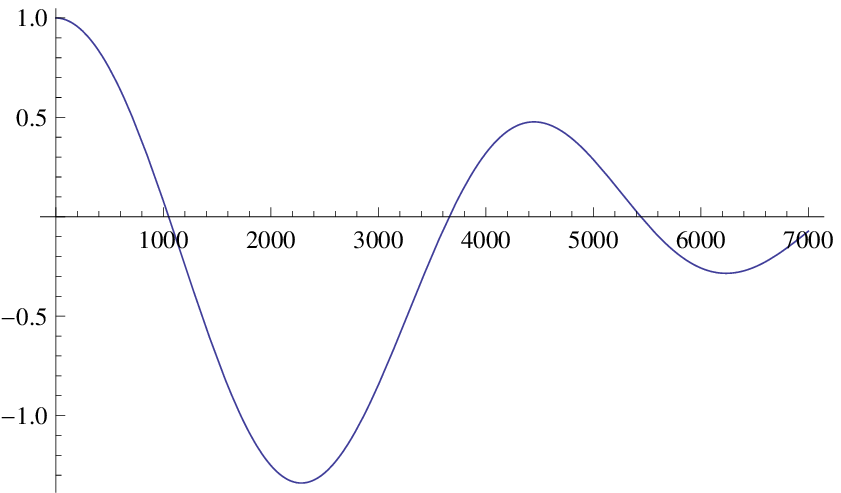}}
%\begin{figure}[ht]
%\epsfig{file=ggbb11.eps,width=350pt, height=150pt}
\\
Fig. 3. Behavior of form-factor $F_V(p)$ for $p\,<\,7000\,GeV$, $G_V$ is defined by (\ref{G}). 
\bigskip
\\
We use form-factor $F_V(p)$ for calculation of cross-sections with the aid of CompHEP package~\cite{Boos}. With value~(\ref{G}) we have for the cross-section at TEVATRON 
for production of $jet\,jet\,(W,\,Z)$~\cite{CDF}  
\begin{eqnarray}
& &\sigma_{jjW,Z}\,\simeq \,1.1\,pb\quad (M_V = 140\,GeV)\,;\label{jjCDF}\\
& &\sigma_{jjW,Z}\,\simeq \,1.2\,pb\quad (M_V = 130\,GeV)\,.\nonumber
\end{eqnarray}
These values do not contradict both CDF~\cite{CDF} ($\sigma = 4.0\pm 1.2\, pb$) and D0~\cite{D0} ($\sigma < 1.9\, pb$) data.

Let us denote these states as $V, V^{\pm}$. Then neutral state V has significant BR for decay $V \to b \bar b,\; BR(b \bar b)\,=\,0.143$~\cite{AZ12}. The cross-section of V production with accompanying $W^\pm$ at TEVATRON also is easily extracted from~\cite{AZ12} results with account of value~(\ref{G}): $\sigma(W^\pm\,V)\,=\,1.3\,pb$. Thus we have
\begin{equation}
 \sigma(W^\pm\,V)\times BR(b \bar b)\simeq 0.17\,pb\,;\label{FLBB}
\end{equation}  
that is to be compared with experimental number~\cite{D0CDF}, which was obtained in the course of the SM Higgs search:
\begin{eqnarray}
& & \sigma(W^\pm\,H)\times BR(b \bar b)\,=\,0.23^{+0.09}_{-0.08}\,pb\,;\label{sigbrb}\\
& & \sigma(W^\pm\,H)\times BR(b \bar b)_{SM}\,=\,0.12\pm 0.01\,pb;\nonumber
\end{eqnarray}
where we also show the SM value for this quantity calculated on assumption of the data being due to the would-be 125.5 GeV Higgs. As a matter of fact, experiment does not contradict both options but agrees the W-vector bound state option~(\ref{FLBB}) rather better.

For comparison with LHC data we calculate also the effect of $jet\,jet$  decay of 135 GeV V state. For $p\,p,\,\sqrt{s} = 7\,TeV$ we have
\begin{equation}
 \sigma_{jjW,Z}\,=\,4.6\,pb\,;\label{LHCjj}
\end{equation}  
that agrees recent data~\cite{LHCjj} $\sigma_{jjW,Z}\,<\,5\,pb\,$.
\section{Comparison to experiments}

Thus we have scalar state $X$ with coupling~(\ref{XWW},\ref{GGX}) and vector state $V^a$ with coupling~(\ref{VWW},\ref{G}). In calculations of decay parameters and cross-sections we use CompHEP package~\cite{Boos}.
Cross-sections of $X$ production at LHC are presented in~(\ref{Section}). Branching ratios see~(\ref{decay}).

From~(\ref{Section}, \ref{decay}) we have for (quite unusual for the Higgs) decay $X \to \gamma l^+ l^-\,(l = e,\,\mu)$ the following value
\begin{eqnarray}
& &\sigma\times BR(X \to \gamma\,l^+\,l^-)_{calc}\,=\,\sigma_{\gamma \gamma\,SM}\mu_{\gamma \gamma\,calc}\times\nonumber\\
& &\frac{BR(X \to \gamma\,l^+l^-)}{BR(X \to \gamma\,\gamma)}\,=\,0.0075\,pb\,.\label{gll}
\end{eqnarray}
This prediction is decisive for checking of the option under discussion.
Remind that we have 
$$
\sigma_{\gamma \gamma}(SM)\,=\,\sigma_H\,BR(H \to \gamma \gamma)\simeq\,0.04\,pb\,.
$$ 
Our value for the same quantity from~(\ref{Section}, \ref{decay}) $\sigma_{\gamma \gamma}\,=\,0.079\,pb$~(\ref{CS0}), that essentially exceeds the SM value $\sigma(SM)$. 
Note that branching ratios~(\ref{decay}) does not depend on the value of $G_X$.
The main results are presented in the following Table 1.
\begin{center}
\begin{tabular}{|c|c|c|c|}
\hline
  & $\mu_{exp}$ & $\mu_{calc}$ &$\mu_{eff}$\\
\hline
$H(X)\to \gamma \gamma$ ATLAS  & $1.8\pm 0.5$ & $1.9$ & -- \\
\hline
$H(X)\to \gamma \gamma$ CMS  & $1.6\pm 0.4$ & $1.9$ & -- \\
\hline
$H(X)\to 4\,l$ ATLAS  & $1.2\pm 0.6$ & $1.05$ & --  \\
\hline
$H(X)\to 4\,l$ CMS  & $0.7\pm 0.4$ & $1.05$ & -- \\
\hline
$H(X)\to b \bar b$ ATLAS  & $0.48^{+2.17}_{-2.12}$ & $0$ & $ 1.01$ \\
\hline
$H(X)\to b \bar b$ CMS  & $0.15^{+0.73}_{-0.66}$ & $0$ & $ 1.01$ \\
\hline
$H(X)\to \tau \bar \tau$ ATLAS  & $0.16^{+1.72}_{-1.84}$ & $0$ & $ 2.5$ \\
\hline
$H(X)\to \tau \bar \tau$ CMS  & $-\,0.14^{+0.76}_{-0.68}$ & $0$ & $2.5$ \\
\hline
$H(X)\to b \bar b$ FNAL  & $1.97^{+0.74}_{-0.73}$ & $0$ & $1.42$ \\
\hline
\end{tabular}\\
\bigskip
Table 1. Comparison of experimental data to SM Higgs option and the W-hadrons option. Column $\mu_{eff}$ deals with effects of production and decay of vector state $V$.
\end{center} 
The last line of Table 1 describes recent joint results of CDF and D0 on detection of $b \bar b$ pair production in region of effective masses  $120\,GeV < M <150\,GeV$~\cite{D0CDF}. This result may be considered as a confirmation of data~\cite{LHC1,LHC2,LHC3,LHC4}. In the framework of the present interpretation we prescribe this effect to production of the resonance $V(140)$. With account of this remark we calculate values of $\chi^2$ per number of degrees of freedom from the Table 1. We have for the two possibilities:
\begin{equation}
\frac{\chi^2_{SM}}{N}\,=\,1.16\,;\quad
\frac{\chi^2_X}{N}\,=\,0.24\,;\quad N\,=\,9\,.\label{chi2}
\end{equation}
The first value corresponds to SM Higgs and the second one corresponds to the option of $W$-hadrons. As a matter of fact both options are compatible with data, however the second one seems for the moment to be preferable.    
Resonance $V(140)$ also give contribution to process $p~+~p~\to (W,Z) + jet\,jet +...$~. CMS result~\cite{LHCjj} gives limitation for possible contribution $\sigma < 5\,pb$ of a resonance with mass $120\,GeV < M_R <150\,GeV$. The contribution for this process of the resonance V(140) is calculated above~(\ref{LHCjj}).
Thus we have here also absence of a contradiction. We would hope that the forthcoming refinement of data should decide definitely for one definite variant\footnote{Of course, one have to bear in mind also other options for interpretation of the effect.}.  For the decisive criterion for the  discrimination of two variants being discussed we would emphasize the importance of channel $X \to \gamma\, l^+ l^-$. For this decay mode from~(\ref{Section}, \ref{decay}) we predict
\begin{equation}
\sigma_X\,BR(X \to \gamma l^+ l^-)\,=\,(0.0075 \pm 15)\,pb;
\label{All1}
\end{equation}
whereas for SM Higgs option such process is negligible. 
The decay~(\ref{All}) gives $N \simeq 70$ events for already achieved luminosity~\cite{LHC1,LHC2,LHC3,LHC4}. This channel might serve for accurate test of our results.

There is also promising process $p + p \to \gamma + X + ...$, with 
cross-section strongly exceeding the cross-section of the process $p + p \to \gamma + H + ...$.  This is due to $X\,Z\,\gamma$ vertex in interaction
~(\ref{XWW}). 

For illustration of effects we show in Table 2 the approximate number of events for processes under discussion. We present 3 values of the total energy: 7~TeV, 8~TeV and 14~TeV.

\begin{center}
\begin{tabular}{|c|c|c|c|}
\hline
$\sqrt{s}\,;\,L$ & $7\,TeV$ & $8\,TeV$ & $14\,TeV$\\
\hline
$N(X\to \gamma \gamma)$ & 380 & 1400 & 5900\\
\hline
$N^{SM}(H\to \gamma \gamma)$ & 200 & 780 & 3300\\
\hline
$N(\gamma + (X \to 2 \gamma))$ & 17.5 & 66 & 285\\
\hline
$N^{SM}(\gamma + (H \to 2\gamma))$ & 0.015 & 0.056 & 0.0243\\
\hline
$N(X \to  \gamma e^+ e^-)$ & 21 & 77 & 322\\
\hline
$N(X \to  \gamma \mu^+ \mu^-)$ & 15 & 53 & 223\\
\hline
$N^{SM}(H \to  \gamma l^+ l^-)$ & 1.2 & 4.5 & 19.3\\
\hline
\end{tabular}
\\
\bigskip
Table 2. Number of events for processes (with 100\% efficiency). Energies $7 GeV,\, 8 GeV,\, 14 GeV$ correspond respectively to luminosities $5\,fb^{-1},\,15\,fb^{-1},\,30\,fb^{-1}$.
\end{center}

We also would draw attention to difference of our predictions with the SM results in decay channel $X\to b \bar b$. For SM Higgs which is usually considered for explanation of would-be $125\,GeV$ state this decay is dominant, whereas our result~(\ref{decay}) gives extremely small $BR\,\simeq 3\,10^{-5}$ (see Table 1). We would emphasize that SM Higgs interpretation could not be considered as proved unless $b \bar b$ channel with the proper intensity would be detected.

We would also draw attention to quite promising process $p\,p\,\to\,\gamma\,+\,X\,+...$ with $X\,\to\, \gamma\,\gamma$. Our option gives for the process cross-section\\ $\sigma(\gamma, X\to\,2\gamma+ ...)\,\simeq\,3.6\,fb$ at LHC, that for already reached luminosity $4.8\,fb^{-1}$ gives around 17 events, whereas for the SM Higgs option the effect is negligible. This process could provide a decisive test of our proposal, the more so as the amount of experimental data will increase in the near future.

\section{Conclusion}

Thus we have an alternative interpretation of LHC phenomenon at $125.5\,GeV$.
The overall data do not contradict both the SM Higgs option and the  description in terms of the scalar W-hadron $X$ with account of the vector W-hadron $V$, which we discuss here. However our estimates of the effects seem to fit data rather better. The forthcoming increasing of the integral luminosity will
undoubtedly discriminate this two options. Especially we would draw attention to processes 
\begin{eqnarray}
& &p\,p\,\to (X\to\gamma\, l^+ l^-)+...;\nonumber\\
& &p\,p\,\to \gamma + (X\to\gamma \gamma)+...;\nonumber
\end{eqnarray}
in which according to Table 2 the effect essentially exceeds the SM 
predictions.

We would draw attention to the non-perturbative effects, which are decisive for the presented option. Just $W$-hadrons in case of confirmation of their existence would follow from non-perturbative electro-weak physics, almost in the same way as the usual hadrons follow from non-perturbative effects in QCD.   

%\section{}
%\label{}



%% References


\bibliographystyle{model1a-num-names}
\bibliography{<your-bib-database>}



\end{document}